\documentclass[twocolumn,prl,showpacs]{revtex4-1}
\usepackage[latin9]{inputenc}
\usepackage{amsmath}
\usepackage{amssymb}
\usepackage{graphicx}

\makeatletter
\usepackage{bm}

\usepackage[]{sidecap}
\sidecaptionvpos{figure}{c}

\makeatother

\begin{document}
\title{Quantum transport through a coupled non-linear exciton-phonon system}
\author{Sima Pouyandeh$^{1,2,3}$, Hadi Z. Olyaei$^{4}$}
\affiliation{$^{1}$Instituto de Telecomunicacoes, Physics of Information and Quantum
Technologies Group, Portugal}
\affiliation{$^{2}$Instituto Superior Tecnico, Universidade de Lisboa, Portugal}
\affiliation{$^{3}$Centre de Biophysique Moleculaire, (CBM), CNRS UPR 4301, Rue
C. Sadron, 45071, Orleans, France}
\affiliation{$^{4}$CeFEMA, Instituto Superior Técnico, Universidade de Lisboa,
Av. Rovisco Pais, 1049-001 Lisboa, Portugal}
\begin{abstract}
Wavepacket transport across a nonlinear region is studied numerically
at zero and finite temperatures. In contrary to the zero temperature
case which demonstrates ballistic transport, finite temperature lattice
vibrations suppresses the transport drastically. The interface between
the linear and the nonlinear chain plays the role of a high barrier
at finite temperature when anharmonicity factor $\beta$ is small
compared to the typical inverse cubic interatomic distance. Inverse
participation ratio of the central region shows that for small anharmonicity
and finite temperatures lattice vibrations give rise to self-trapping
in the nonlinear chain which lasts for considerable times with a subdiffusive
leakage of the wavepacket, almost equally, to both leads. The scenario
changes when the anharmonicity becomes comparable with average inverse
cubic interatomic distances as the lattice dynamics gives a profound
boost to the transmission and starts to be almost transparent for
the incoming pulse.
\end{abstract}
\maketitle

\section{Introduction}

Nonlinear lattice dynamics is one of the pivotal subjects of modern
condensed matter physics and nonlinear science. There has been enormous
activity since the pioneering works of Fermi, Pasta, Ulam and Tsingou \citep{Campbell2005,Gallavotti}
for many years in the study of the spreading of wave packets known
as energy excitations in numerous nonlinear models including the Fermi-Pasta-Ulam(FPU)
model \citep{Bourbonnais1990,Zavt1993,Leitner2001,Snyder2006} and
the discrete nonlinear Schrödinger equation(DNLSE)\citep{Shepelyansky1993,Molina1998,Pikovsky2008}.
Many experiments were performed within the context of wave packets
spreading in nonlinear lattices with ultracold atomic condensates
\citep{Clement2006,Sanchez-Palencia2007,Billy2008,Lucioni2011}, light
energy transport in light-harvesting biomolecules and photosynthesis
\citep{VanAmerongen2000,Blankenship2002,Engel2007,Read2009,Collini2010,Romero2014}.

Among various nonlinear models, DNLS effectively describes the dynamics
of excitons interacting with lattice vibrations dynamics. The presence
of nonlinearity in the DNLS keeps track of some interesting features
in Bose-Einstein condensates BECs, \citep{Xue2008,Wang2006} optical
lattices, \citep{Akhmediev2009,Ponomarenko2006}. Moreover, energy
Localization or propagation of the stable solutions of DNLS, such
as breathers and solitons, is associated to energy transfer in biological
systems. \citep{Braun1998,Hennig1999}. There have been several investigations
on the spreading of initially localized wave packets in a lattice
with vibrations induced nonlinearity\citep{Shepelyansky1993,Kopidakis2008,Pikovsky2008,Flach2009,Skokos2009,Mulansky2010,Laptyeva2010,Iomin2010,Larcher2009,Iubini2015}.
The effect of nonlinearity on the transport features of disordered
dynamical systems are also studied at length\citep{Mulansky2010,Kopidakis2008,Flach2010,Garcia-Mata2009,Mulansky2013,Ivanchenko2011}.
Vibrations induced nonlinearity gives rise to terms with cubic non-linearity
in the DNLS equation which can be understood as an adiabatic-type
approximation of phonon degrees of freedom\citep{Chen1993,Molina1993}.

In this article,  the adiabatic assumption is lifted by treating the
lattice vibrations as a nonlinear FPU coupled with the Schrodinger
equation governing the dynamics of the propagating wave packet. Hence,
from the point of view of the expanding wave packet, the underlying
lattice is a classical object which modifies the nearest neighbor
hopping integrals and onsite potentials in the usual tight-binding
approximation. The interplay between the quantum part(exciton) and
the classical lattice, simulated as  FPU chain, can significantly
change electronic transport in the system. Studies on the coupled
dynamical nature of exciton and underlying protein vibrations signify
the role of vibrations modes in exciton energy transfer \citep{OReilly2014,Tiwari2013,Mennucci2011,Plenio2008,Lee2007,Wang2007,Adolphs2006,Pouyandeh2017a,Novo2016}.
\textcompwordmark Besides the fact that our grasp of energy spreading
phenomenon in nonlinear lattices has enhanced over decades, there
is still a universal mechanism missing in understanding the effect
of out of equilibrium leads attached to the system is still open.

In this paper, we study energy transport through a nonlinear chain
connected to two semi-infinite leads at its endpoints. The leads are
regular linear one-dimensional lattices simulating the reflectionless
contacts. The setup is of particular relevance for nano-bio-electronic
devices. The central interest is to study the general features of
energy/charge transport in such a device and to explore how an incident
pulse travels through the nonlinear semiclassical lattice from a reservoir
towards the central region. The configuration present in this paper
takes also the linear/nonlinear interfaces into consideration as most
semiconducting electronic devices have metallic tails with a minimum
resistance and energy dissipation. So the linear/nonlinear interface
plays a role of a dynamically varying barrier which needs to be taken
into account.

In Sec. \ref{sec:Model-and-methods} we introduce the model describing
the physical setup and the relevant tight-binding formulation along
with the numerical method used to obtain the transmission. In Sec.
\ref{sec:Results-and-Discussion}, includes some representative results
of transmission and inverse participation ratio. In Sec. \ref{sec:Conclusion}
We conclude with  a short summary. Sec. \ref{sec:Conclusion} contains
the details of obtaining semiclassical exciton-phonon coupling equations.

\section{Model and methods\label{sec:Model-and-methods}}

Our system consists of two linear 1D chains serving as semi-infinite
leads and a central region consisting of a finite length chain coupled
to the underlying 1D classical lattice. The aim is to send a pulse
from the left lead to the right to pass the non-linear chain as a
barrier and study the transmission and reflection and other properties
of this transport to get an insight for the behavior of this energy
transport. In light-harvesting systems this pulse is, in fact, the
wave function of an exciton, excited from a two-level quantum system
of a pigment in a photosynthesis structure or a charge particle transporting
in theses light-harvesting complexes. In this regard, we can investigate
the role of non-linearity of the inter-atomic potentials underlying
lattice as well as the role of local and non-local dynamical disorder
in these complexes. The propagation of the pulse which is sent into
the lattice and inherits a quantum nature together with the vibrations
of the lattice are governed by the general following Hamiltonian

\begin{eqnarray}
H & = & \sum_{_{mn}}J_{mn}\left(u_{m},u_{n}\right)B_{m}^{\dagger}B_{n}+\sum_{n}\frac{P_{n}^{2}}{2M}+\sum_{mn}V\left(u_{m},u_{n}\right)\nonumber \\
\label{eq:H}
\end{eqnarray}

Where $M$ is the bead mass, $B_{n}^{\dagger}$,$B_{n}$ are creation
and annihilation operators on site $n$ respectively and $u_{n}$
is the displacement of $n$th bead with respect to its equilibrium
position. In the Holstein model, the interaction of the quantum system
with the lattice is defined by $J_{mm}=\epsilon_{m}+\chi u_{m}$.
In this model all the off-diagonal elements of $J_{mn}$ are set to
zero except for the nearest neighbors ($J_{m,m+1}=J_{m+1,m}=1$).
In the so-called SSH (Su-Schrieffer-Heeger) model which is used mostly
for the organic semiconductors the couplings are $J_{m,m+1}=J-\chi\left(u_{m}-u_{m+1}\right)$
and $J_{mm}=\epsilon_{m}$ \citep{Su1979}.

Here, we use a more physically realistic model which combines the
two models. The Hamiltonian of an exciton propagating in the lattice
is given by the tight-binding Hamiltonian

\begin{eqnarray}
H_{e} & = & \sum_{n=1}^{L}\epsilon_{n}\left(u\right)B_{n}^{\dagger}B_{n}+\sum_{n=1}^{L}J_{n}\left(u\right)\left(B_{n+1}^{\dagger}B_{n}+B_{n}^{\dagger}B_{n+1}\right)\nonumber \\
\label{eq:He}
\end{eqnarray}

In which the effective interactions are modulated as following \citep{Iubini2015}

\begin{eqnarray}
\epsilon_{n} & = & \epsilon_{0}+\chi_{E}\left(u_{n+1}-u_{n-1}\right)\label{eq:En}\\
 &  & ,\nonumber \\
J_{n} & = & J_{0}+\chi_{J}\left(u_{n+1}-u_{n}\right)\label{eq:Jn}\\
\nonumber
\end{eqnarray}

In the above equations $\epsilon_{0},J_{0}$ refer to the unperturbed
values of site energies and hopping integrals respectively. Also,
the parameters $\chi_{E},\chi_{J}$ are the strength of the exciton-lattice
couplings.

Moreover, the Hamiltonian of the lattice which includes nonlinear
interactions between nearest neighbor oscillators is given by

\begin{eqnarray}
H_{l} & = & \sum_{n=0}^{L}\frac{p_{n}^{2}}{2M}+\frac{\kappa}{2}\left(u_{n+1}-u_{n}\right)^{2}+\frac{\beta}{4}\left(u_{n+1}-u_{n}\right)^{4}\label{eq:Hl}\\
\nonumber
\end{eqnarray}

Where, $\beta$ demonstrates the non-linearity of the lattice. In
order to determine the evolution of the system, we obtain the equations
of motion (EOM) for both quantum and classical parts. By defining
a trial wave function for a single-exciton manifold $|\psi\left(t\right)\rangle=\sum_{n}b_{n}\left(t\right)B_{n}^{\dagger}|0\rangle$
and using it in the Schrodinger equation ($i\hbar\frac{\partial}{\partial t}|\psi\left(t\right)\rangle=H|\psi\left(t\right)\rangle$)
together with some identity relations we get the equation of motion
of the exciton

\begin{eqnarray}
i\hbar\frac{\partial}{\partial t}b_{j}\left(t\right) & = & \epsilon_{j}\left(u\right)b_{j}\left(t\right)+J_{j-1}\left(u\right)b_{j-1}\left(t\right)+J_{j}\left(u\right)b_{j+1}\left(t\right)\nonumber \\
\label{eq:EOM-ex}
\end{eqnarray}

The EOM of the lattice will be that of a set of coupled oscillators
driven by the exciton wave-function. From Newton's law ($M\ddot{u}_{n}=-\frac{\partial H}{\partial u_{n}}$),
we can write:

\begin{eqnarray}
M\ddot{u}_{n} & = & -\frac{\partial}{\partial u_{n}}\langle\psi|H|\psi\rangle\nonumber \\
 & = & -\frac{\partial}{\partial u_{n}}\langle\psi|H_{e}+H_{l}|\psi\rangle\label{eq:newton}\\
 & = & F_{n}^{e}+F_{n}^{l}\nonumber \\
\nonumber
\end{eqnarray}

With $H_{e}$ and $H_{l}$ given in \ref{eq:} and \ref{eq:Hl} respectively.

Using Davydov trial wave function defined before, identity relations
and normalization condition, after a few calculations similar to excitons,
we obtain the EOM for the phonons as the classical part of the system

\begin{eqnarray}
M\ddot{u}_{n} & = & \kappa\left(u_{n+1}+u_{n-1}-2u_{n}\right)\nonumber \\
 & + & \beta\left[\left(u_{n-1}-u_{n}\right)^{3}+\left(u_{n+1}-u_{n}\right)^{3}\right]\nonumber \\
 & + & \chi_{E}\left(b_{n+1}^{*}\left(t\right)b_{n+1}\left(t\right)-b_{n-1}^{*}\left(t\right)b_{n-1}\left(t\right)\right)\nonumber \\
 & + & 2\chi_{J}Re\left(b_{n+1}^{*}\left(t\right)b_{n}\left(t\right)-b_{n-1}^{*}\left(t\right)b_{n}\left(t\right)\right)\label{eq:EOM-ph}\\
\nonumber
\end{eqnarray}

This equation together with the EOM of the exciton (Eq.\ref{eq:EOM-ex})
are integrated simultaneously by a fourth order Runge-Kutta method
to determine the evolution of the coupled exciton-phonon non-linear
system.

\subsection{Initial and Boundary conditions}

In order to investigate the energy transport in the system, we formulate
the problem in terms of a localized wave packet moving towards a non-linear
scattering region to get a motion picture displaying of reflection
and transmission. We represent the initial state of the exciton by
a Gaussian wave packet\citep{Goldberg1967}

\begin{align}
\psi(x,0) & =e^{ik_{0}x}e^{-(x-x_{0})^{2}/2\sigma_{0}^{2}}\label{eq:wave packet}\\
\nonumber
\end{align}

This packet is centered around $x=x_{0}$ with a spread in $x$ dependent
on $\sigma_{0}$. The factor $e^{ik_{0}x}$ makes the wave function
move to the right with average momentum $k_{0}$. The choices of $x_{0}$
and $\sigma_{0}$ are governed by the boundary conditions and two
more restrictions in the general situations which must be imposed
in order that box normalization not give rise to difficulties. First
of all, the packet is not allowed to travel so far that it hits the
walls \citep{Goldberg1967}. For the total length of the system $L$
(including both leads and the non-linear chain), this restriction
can be ensured by letting the center of the packet start from $L_{1}=x_{0}$,
move no farther to the right than $L_{2}$, provided that the length
of two leads is much larger than the chain. This can be accomplished
by requiring that the average velocity of the wave packet be

\begin{align}
k_{0} & \approx v_{0}=\frac{L_{2}-L_{1}}{T}\label{eq:restriction1}\\
\nonumber
\end{align}

Where $T$ is the total time that is taken by the packet to move from
$L_{1}$ to $L_{2}$. In fact, the average momentum $k_{0}$ should
be chosen in a way that the wave packet is transmitted through the
chain fast enough without delocalization over the lead.

The second of the two restrictions concerns the fact that the reflected
and transmitted packets must continue to be well enough localized
so that at the end of the event they are out of the region of the
potential and still far from the walls. It can be shown that if $\sigma_{0}$
is the initial spread in free space, then after a time $T$ the spread
is given by\citep{Goldberg1967}

\begin{align}
\sigma^{2} & =\left(\sigma_{0}^{4}+4T^{2}\right)^{1/2}\label{eq:restriction2}\\
\nonumber
\end{align}

Here we apply the same criteria for the spreading in the presence
of a potential. For negligible spreading we must have $4T^{2}$ reasonably
small compared with $\sigma_{0}^{4}$.

However, despite these restrictions, as choosing long leads is time-consuming
in terms of the numeric simulations, we have chosen another strategy
to avoid the difficulties caused by wave packet when hitting the walls.
Here we have applied absorbing boundary conditions at the walls in
a way that we can make sure there will be no reflection from the walls
so, the transmission that we calculate corresponds only to the initial
pulse we had and not the reflected waves.

For the boundaries between leads and non-linear chain, we have chosen
fixed boundary conditions which means that at the place where excitation
hits scattering region the wave function should be zero at all times

\begin{align}
 & \psi\left(N_{L}^{left},t\right)=\psi\left(N_{L}^{right},t\right)=0\label{eq:boundaries1}
\end{align}

Where $N_{L}^{left}$ ,$N_{L}^{right}$ are the index of sites at
right and left boundaries of the left and right leads with the lattice,
respectively. This, of course, is applied to initial boundary conditions
at $t=0$ for the same sites as well.

\begin{figure}
\begin{centering}
\includegraphics[width=1\columnwidth]{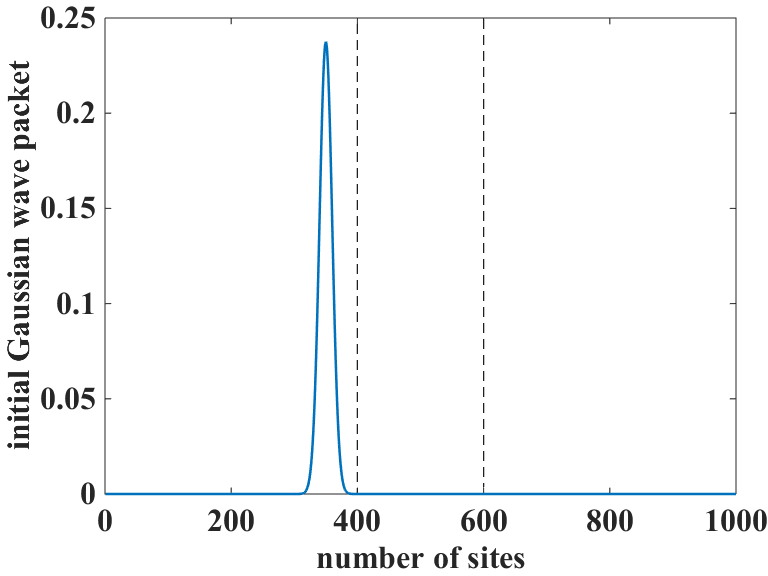}
\par\end{centering}
\caption{Initial pulse localized at $x_{0}=N_{L}-50$ for a system of total
length N=1000. The boundaries of leads and non-linear chain are specified
by dashed lines. All the parameters including energies and couplings
as well as non-linearity are kept one here. \label{fig:Initial-pulse-localized}}
\end{figure}

Figure \ref{fig:Initial-pulse-localized} shows the exciton wave packet
initially localized at $x_{0}$ for a system of total length 1000
with a non-linear lattice of 200 sites between two leads of length
400 each.We have chosen $x_{0}=350$ for the  initial position of
the pulse and $\sigma_{0}=10$ for spreading parameter which is $1\%$
of the total system length.

Regarding the initial conditions for non-linear lattice, we suppose
a characteristic configuration of the variables $u_{n}$ and $p_{n}$
representative of a finite temperature $T$. This can be done by thermalizing
the lattice via a Langevin heat bath at temperature $T$ for a sufficiently
long transient time. For this, we implement a suitable friction term
and a stochastic force to the free lattice EOMs \citep{Iubini2015,Lepri2003}

\begin{align}
M\ddot{u}_{n} & =F_{n}^{l}-\gamma p_{n}+\sqrt{2\gamma T}\xi_{n}\left(t\right)\label{eq:Langevin1}\\
\nonumber
\end{align}

Where $\gamma$ is the coupling strength of the heat bath and $\xi_{n}\left(t\right)$
is a Gaussian white noise with correlation $\langle x_{i}\left(t\right)x_{i}\left(t^{\prime}\right)\rangle=2T\delta\left(t-t^{\prime}\right)$.
We start with a Maxwell distribution at temperature T for lattice
configuration and then perturb momenta with random kicks extracted
from a Gaussian distribution with $0$ mean and $2T\delta t$ variance.
After the transient time $t_{0}$ we disconnect Langevin reservoir
and integrate the equations of motion of the coupled system in order
to study its transport properties.

\section{Results and Discussion\label{sec:Results-and-Discussion}}

\begin{figure}
\begin{centering}
\includegraphics[width=1\columnwidth]{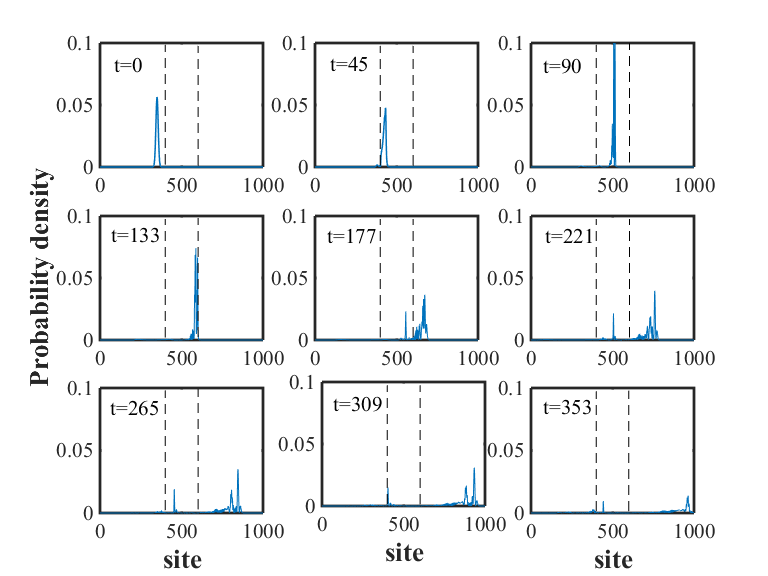}
\par\end{centering}
\caption{Evolution of probability density $|\psi(x,t)|^{2}$ for $x_{0}=350$,
$\sigma_{0}=10$, $N_{c}=200$, $N=1000$ and $t_{f}=40$0 at zero
temperature. All the other parameters including energies, couplings,
non-linearity are kept one here. \label{fig:Evolution-of-probability density}}
\end{figure}

In this section, we study the process of energy transport through
the chain. We study the exciton transfer in different regimes in terms
of temperature, coupling strength and non-linearity. For that, we
integrate the EOMs for both exciton and phonons (Eq.\ref{eq:EOM-ex}
and Eq.\ref{eq:EOM-ph}) simultaneously to treat the reflection-transmission
event in a physical way. Later, we use our model to investigate the
energy transport in an Organic semiconductor as a potentially natural
light-harvesting system. Figure \ref{fig:Evolution-of-probability density}
shows the evolution of the initial excitonic pulse localized at $x_{0}=350$
over the whole system of total length $N=1000$, with two leads of
$N_{lead}=400$ each in the total time $t=40$0 at zero temperature.
The initial width of pulse is $\sigma_{0}=10$ while all the other
parameters including energies, couplings and non-linearity are kept
one here.

We can see that in the case of zero temperature most part of the initial
wave function is transferred through the non-linear chain. In the
next part we will calculate the transmission and study its properties
for different temperatures, non-linearity and coupling strengths.

\subsection{Energy Transport }

To gain insight into the process of energy transport, we study transmission
and reflection phenomena and their relevance to other parameters of
the system. Also to investigate the spread or localization of the
exciton over lattice sites and get information in more details about
its transfer, we obtain the participation ratio during different stages
of the motion. The reflectivity and transmission probabilities are
determined by recording and integrating the reflected and transmitted
probability currents at $x=x_{r}$ and $x=x_{t}$, two single positions
before and beyond the scattering region, respectively. The integration
must be performed over a duration spanning the entire scattering event
(that is, waiting until the probability current has vanished at the
single point). The transmission is \citep{Dimeo2014}

\begin{align}
T & =\int_{0}^{t_{final}}dt^{\prime}j\left(x_{t},t^{\prime}\right)\label{eq:transmission1}\\
\nonumber
\end{align}

Where, $j(x_{t},t^{\prime})$ is the probability current given by
the usual definition

\begin{align}
j(x,t) & =\frac{1}{2i}\left[\psi^{\star}\frac{\partial\psi}{\partial x}-\frac{\partial\psi^{\star}}{\partial x}\psi\right]\label{eq:probabilty current1}\\
\nonumber
\end{align}

Similar to Eq. (\ref{eq:transmission1}), reflection can be obtained,
using

\begin{equation}
R=-\int_{0}^{t_{final}}dt^{\prime}j\left(x_{r},t^{\prime}\right)
\end{equation}

Or, we can simply use Eq.(\ref{eq:transmission1}) and calculate the
reflection.

In our model of discrete lead and lattice sites, we have used the
following probability current in discrete case

\begin{eqnarray}
\jmath & = & i\sum_{i=1}^{N}J_{i}\left(b_{i}^{\star}b_{i+1}-b_{i+1}^{\star}b_{i}\right)\label{eq:net flux}\\
\nonumber
\end{eqnarray}

It is obtained directly from the quantum current flow $\jmath_{i}\text{=}i\left[H_{e},\hat{N_{i}}\right],\hbar=1$
for excitons, using identity relations and doing a bit of algebra.

\begin{figure}
\begin{centering}
\includegraphics[width=1\columnwidth]{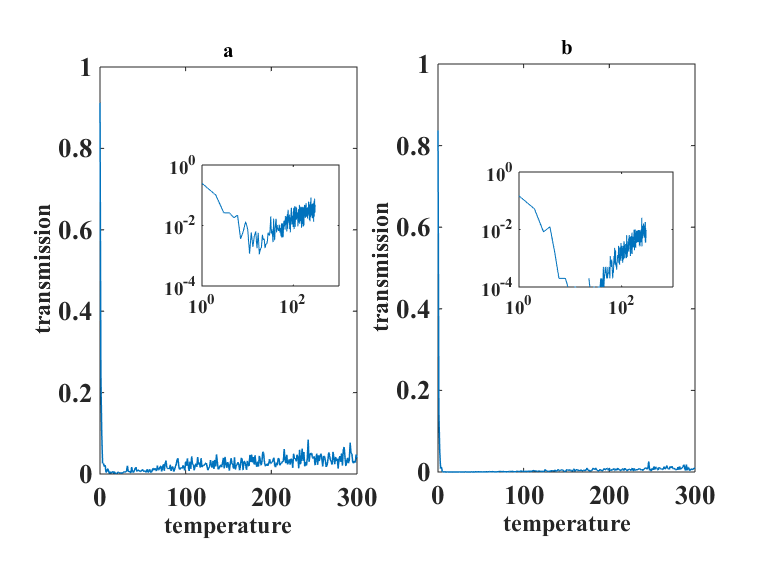}
\par\end{centering}
\caption{Transmission versus temperature for a system of total length $N=900$
with $N_{c}=100$ over transient time $t=700$ at $x_{t}=505$ in
left and a system of total length $N=1000$ with $N_{c}=200$ over
transient time $t=700$ at $x_{t}=605$ in right panel. The length
of leads in both systems is $N_{lead}=400$ each and the initial width
of the pulse is $\sigma_{0}=10$. All the other parameters including
energies, couplings and non-linearity are kept one. \label{fig:Trans_Temp_L100-200}}
\end{figure}

In figure \ref{fig:Trans_Temp_L100-200}, the transmission is plotted
in terms of temperature for two systems of total length $N=900$ and
$N=1000$ with $N_{c}=100$ , $N_{c}=200$ over total transient time
$t=700$ and $t=1000$ respectively. The length of leads in both systems
is $N_{lead}=400$ each and the initial width of the pulse is $\sigma_{0}=10$.
All the other parameters including energies, couplings and non-linearity
are kept one. The inset shows the logarithmic scale of the transmission
over temperature for this system.

It is clear from the figures that by increasing temperature from zero,
the transmission drops and then oscillates around small amounts. Here
non-linear chain acts as an insulator which absorbs the energy of
pulse, by reflecting and re-transmitting this energy from its boundaries
several times inside the scattering region. However it seems that
transport is slightly enhanced by increasing the temperature. But,
in fact, it is due to the non-linearity properties of the system that
in higher temperature lead to increase the amplitude of the osculations
in the non-linear term and so slightly raise the average transport.

\begin{figure}
\begin{centering}
\includegraphics[width=1\columnwidth]{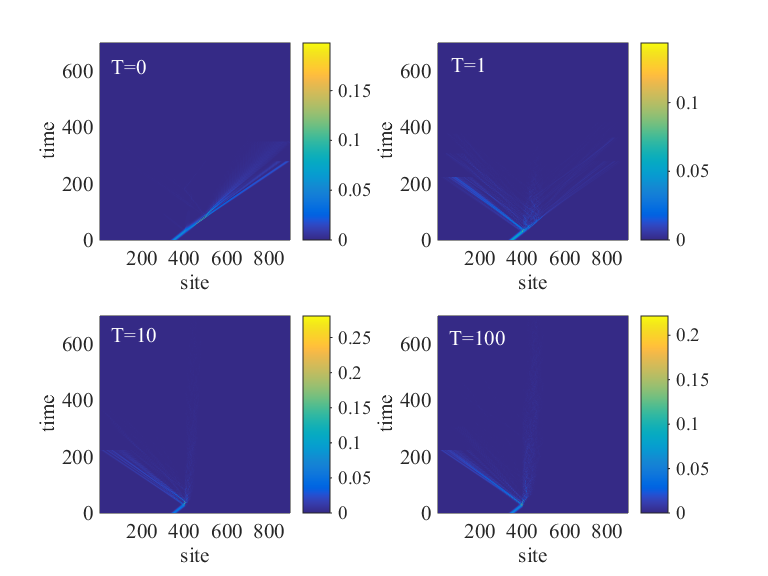}
\par\end{centering}
\caption{Density plot of evolution of $|b|^{2}$ over a system of total length
$N=900$ with $N_{c}=100$ and $N_{lead}=40$0 for temperatures $T=0,1,10,100$.\label{fig:dp_pulse_L900}}
\end{figure}

\begin{figure}
\begin{centering}
\includegraphics[width=1\columnwidth]{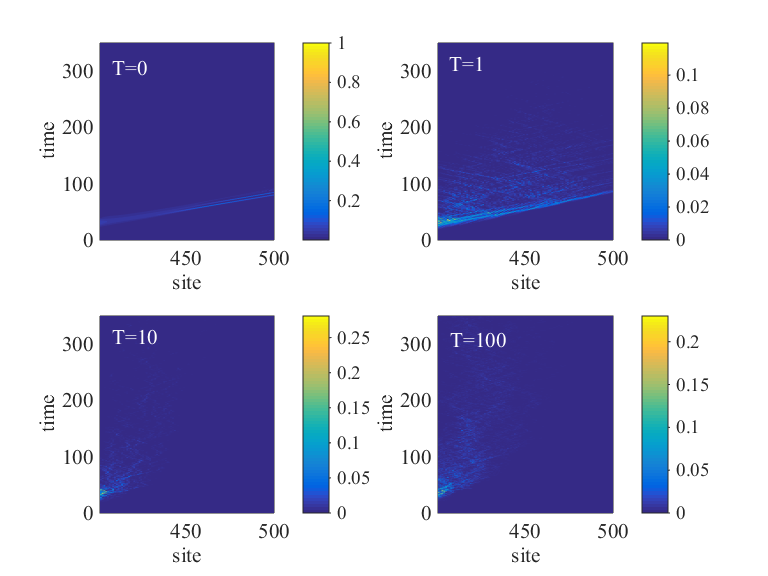}
\par\end{centering}
\caption{Density plot of evolution of $|b|^{2}$ over the non-linear sites
of a system of total length $N=900$ with $N_{c}=100$ and $N_{lead}=40$0
at temperatures $T=0,1,10,100$..\label{fig:dp_pulse_L10}}
\end{figure}

Figure \ref{fig:dp_pulse_L900} shows the density plot of wave function
amplitude $|b|^{2}$ of the initial pulse during the evolution over
all lattice sites of a system of total length $N=900$ with non-linear
chain of length $N_{c}=100$ and two leads of length $N_{lead}=40$0
each for different temperatures. Figure \ref{fig:dp_pulse_L10} is
density plot of pulse amplitude evolution only over non-linear lattice
sites for the same temperatures as figure \ref{fig:dp_pulse_L900}.
As it is seen in the presence of temperature the pulse starts to spread
over scattering region. In fact, at the higher temperatures the pulse
even hardly reaches the second boundary and the transmission is almost
zero.\\

To have a better insight into what happens inside this scattering
region we have computed the inverse participation ratio (IPR) for
different temperatures. IPR is a measure of localization and is defined
as the average of the absolute value of the fourth power of the wave
function. The maximum value of this quantity which is $IPR_{max}=1$
with a choice of normalization is reached when the wave function is
completely localized while the minimum value is reached for a perfectly
uniform state.

\begin{figure}
\begin{centering}
\includegraphics[width=1\columnwidth]{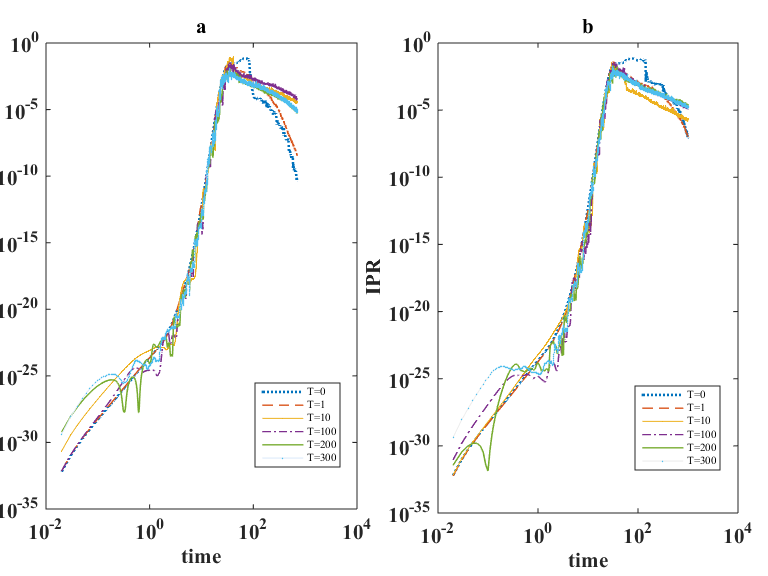}
\par\end{centering}
\caption{Evolution of IPR in non-linear chain at different temperatures for
left figure: $N=900$, $N_{c}=100$, $t=700$ and right figure: $N=1000$,
$N_{c}=200$, $t=1000$ . \label{fig: IPR_L100_L200_T0_300}}
\end{figure}

\begin{figure}
\begin{centering}
\includegraphics[width=1\columnwidth]{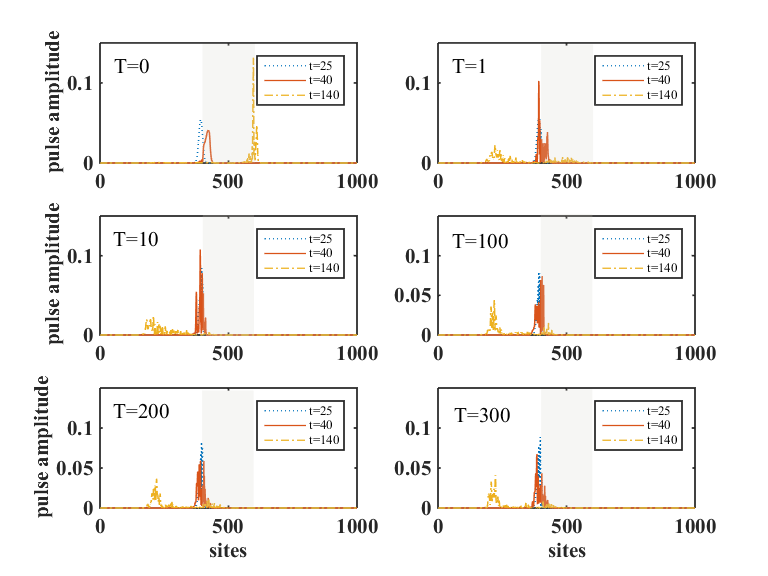}
\par\end{centering}
\caption{$|b|^{2}$ over lattice sites of a system of total length $N=1000$
with non-linear chain of $N_{c}=200$ and two leads of length $N_{lead}=400$
each for different temperature at three different times $t=25$, when
pulse reaches non-linear region, $t=40$ the time of maximum IPR and
$t=140$ the time when wave reaches second boundary. \label{fig: b2_L200_T0_300}}
\end{figure}

In figure \ref{fig: IPR_L100_L200_T0_300} the evolution of IPR in
the non-linear region is plotted at different temperatures for two
systems of total length $900$ and $1000$ with $N_{c}=100$ , $N_{c}=200$
respectively while the length of each two leads in each system is
$400$. What is understood from these figures is that when the wave
function reaches the scattering region, IPR goes up for a short time
which means pulse is entering the region. However, this local wave
function cannot keep its localization and spreads over lattice very
fast. It is obvious that before the time that wave function reaches
the region, IPR should be zero, however after entering the region,
except for zero temperature it cannot keep higher values and before
reaching the second boundary of non-linear region drops to zero which
means that has lost its localization. Figure \ref{fig: b2_L200_T0_300}
shows the screenshot of the wave function amplitude over all lattice
sites of the system with length 1000 for different temperatures at
three times of entering the non-linear region, reaching the second
boundary of this region and a time between these two times close to
average maximum IPR.

\begin{figure}
\begin{centering}
\includegraphics[width=1\columnwidth]{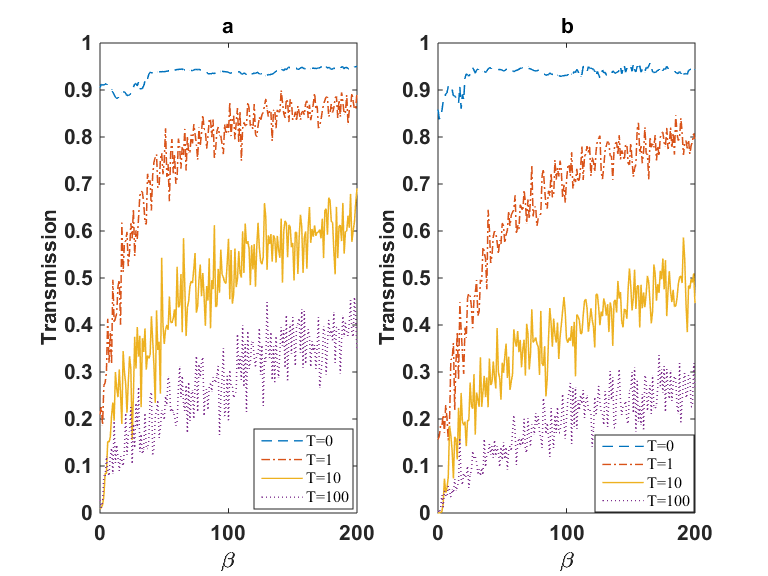}
\par\end{centering}
\caption{transmission versus $\beta$ at different temperatures for $N=900$,
$N_{c}=100$, $t=700$ and $N=1000$, $N_{c}=200$, $t=1000$ in left
and right figures respectively. The length of each lead is $N_{lead}=400$
in both cases. \label{fig:trans-beta}}
\end{figure}

However and despite this inefficient transport in the presence of
temperature, surprisingly by increasing the non-linearity, transmission
goes up. In figure \ref{fig:trans-beta} $a)$ and $b)$ we have plotted
transmission in terms of the non-linearity parameter $\beta$ at three
different temperatures for two systems of total length $900$, $1000$
with non-linear chain of length $100$ , $200$ during the time evolution
of $t=700$ and $t=1000$ respectively. As seen in this figure by
increasing $\beta$, transmission also increases and can reach up
to $90\%$ for $T=1$.

One explanation for this behavior could be hidden in the origin of
coherence in the quantum transport systems such as photosynthetic
complexes. In fact, it is experimentally and theoretically verified
that the coherent exciton-vibrational (vibronic) coupling is of the
origin of long-lasting coherence in the light-harvesting complexes
both natural and artificial. Here, it can be concluded that the non-linearity
is also another parameter of the origin of coherence among other parameters.
However, what is seen here is that, non-linearity works as a catalyst
for the whole system to improve quantum transport. In general, in
a non-linear system there are two types of stable and unstable modes
which can promote or demote the energy transport of the system. The
normal modes will work as a repeater by creating recurrence and memory
preserving of the modes while the unstable modes driven by instabilities
caused by temperature will perturb the system in a destructive way.
In other words, temperature and non-linearity play as rivals to prevent
or promote the energy transport in the non-linear system. At lower
temperatures the positive impact of non-linearity is more profound
and as it is seen in figure \ref{fig:trans-beta} it can help transport
to enhance up to 90 percent.

\begin{figure}
\begin{centering}
\includegraphics[width=1\columnwidth]{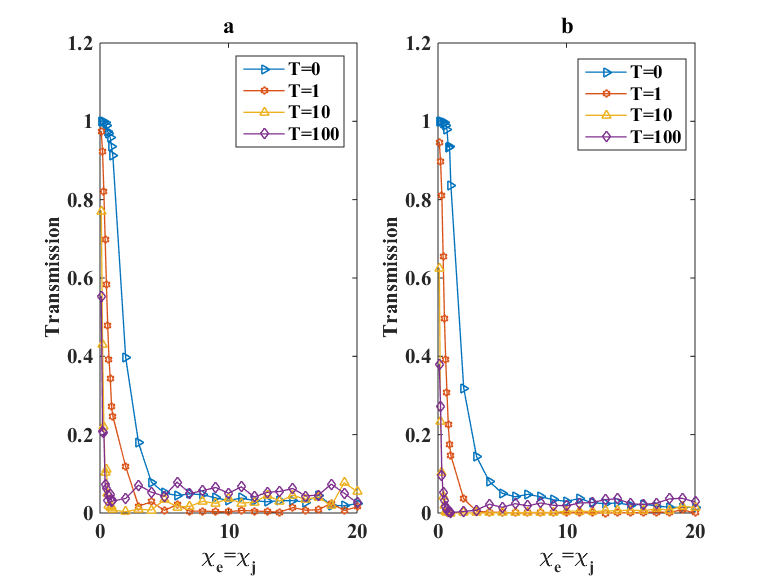}
\par\end{centering}
\caption{Transmission in terms of coupling strength $\chi(\chi_{e}=\chi_{j})$
at different temperatures for $N=900$, $N_{c}=100$, $t=700$ and
$N=1000$, $N_{c}=200$, $t=1000$ in left and right panels respectively.
The length of each lead is $N_{lead}=400$ in both cases. \label{fig:transmission-x}}
\end{figure}

In figure \ref{fig:transmission-x} the effect of couplings between
excitonic (pulse) and vibronic systems on transmission is also shown
for the two systems of length $900$, $1000$ with non-linear chain
of length $100$ , $200$ during the time evolution of $t=700$ and
$t=1000$ respectively. As expected, the most coupled the exciton
is to the vibrations, the less it is free to move and so, the less
the transmission is.

So, we can say that in one hand side, there are temperature and coupling
strengths and on the other hand, non-linearity that interplay like
rivals to help or harm the energy transport in the coupled exciton-phonon
systems.

\section{Conclusion\label{sec:Conclusion}}

We numerically studied the quantum transport of the wavepacket across
a nonlinear FPU chain. The results show that for small anharmonicity
factors, finite temperature acts as an adverse control parameter in
the transmission and causes self-trapping in the central region. Self-trapped
wave leaks to the leads in a subdiffusive manner. Increasing the anharmonicity
enhances transport significantly.
\begin{acknowledgments}
S.P thanks the support from Fundaç$\tilde{a}$o para a Ci$\hat{e}$ncia
e a Tecnologia (Portugal), namely through programme POCH and projects
UID/EEA/50008/2013 and UID/EEA/50008/2019, as well as from the EU
FP7 project PAPETS (GA 323901). Furthermore, H.Z.O and S.P acknowledge
the support from the DP-PMI and FCT (Portugal) through scholarship
PD/ BD/113649/2015 and PD/ BD/52549/2014 respectively. The authors
kindly thank Yasser Omar, Francesco Piazza and Rui A. P. Perdigão for
useful discussions related to this work.
\end{acknowledgments}

\bibliographystyle{apsrev4-1}
\addcontentsline{toc}{section}{\refname}\bibliography{transport_through_nonlinear_chain}

\end{document}